\documentclass[aps, prb, preprint, superscriptaddress,showpacs]{revtex4}

\usepackage{natbib}

\usepackage{graphicx}
\usepackage{dcolumn}
\usepackage{bm}
\usepackage{upgreek}
\usepackage{amsmath}
\usepackage{amssymb}
\usepackage{tabularx}
\usepackage{array}

\newcommand*{\RMn}{$R\text{Mn}_2\text{O}_5$}
\newcommand*{\GdMn}{$\text{GdMn}_2\text{O}_5$}

\begin{document}


\title{Magnetoelectric Phase Diagrams of Multiferroic \GdMn{}}

\author{S. H. Bukhari}
\affiliation{Department of Physics, Bahauddin Zakariya University, Multan 60800, Pakistan}
\affiliation{Institute of Solid State Physics, Vienna University of Technology, 1040 Vienna, Austria}
\author{Th. Kain}
\author{M. Schiebl}
\author{A. Shuvaev}
\author{Anna Pimenov}
\affiliation{Institute of Solid State Physics, Vienna University of Technology, 1040 Vienna, Austria}
\author{A. M. Kuzmenko}
\affiliation{Prokhorov General Physics Institute, Russian Academy of
Sciences, 119991 Moscow, Russia}
\author{X. Wang}
\affiliation{University of Science and Technology, Beijing, China}
\affiliation{Rutgers Center for Emergent Materials and Department of Physics and Astronomy, Rutgers University, New Jersey 08854, USA}
\author{S.-W.Cheong}
\affiliation{Rutgers Center for Emergent Materials and Department of Physics and Astronomy, Rutgers University, New Jersey 08854, USA}
\author{J. Ahmad}
\affiliation{Department of Physics, Bahauddin Zakariya University, Multan 60800, Pakistan}
\author{A. Pimenov}
\affiliation{Institute of Solid State Physics, Vienna University of Technology, 1040 Vienna, Austria}

\date{\today}

\begin{abstract}
Electric and magnetic properties of multiferroic \GdMn{} in external magnetic fields were investigated to map out the magnetoelectric phases in this material. Due to strong magnetoelectric coupling, the dielectric permittivity is highly sensitive to phase boundaries in \GdMn, which allowed to construct the field-temperature phase diagrams. Several phase transitions are observed which are strongly field-dependent with respect to field orientation and strength. The phase diagram for a magnetic field along the crystallographic $a$-axis corresponds well to a polarization step, as induced by 90$^{\circ}$ rotation of Gd magnetic moments. Our results support the model of two ferroelectric sublattices, Mn-Mn and Gd-Mn with strong $R$-Mn ($4f$-$3d$) interaction for the polarization in \RMn{}.

\end{abstract}


\maketitle


\section{\label{intro}Introduction}

Multiferroics are materials in which magnetic and electric orders coexist and which are promising for the development of novel functional materials and devices~\cite{fiebig_jpd_2005, eerenstein_nature_2006, tokura_jmmm_2007, ramesh_nmat_2007}. The coupling between the magnetic and electric degrees of freedom has the consequence that multiferroics exhibit new physical effects: electric and magnetic orders can be modified by external magnetic or electric fields and the propagation of electromagnetic radiation can be controlled.

Recently, the observation of large polarization~\cite{lee_prl_2013} in \GdMn{} and of colossal magnetoelectric effect in $\text{TbMn}_2\text{O}_5$ and $\text{DyMn}_2\text{O}_5$ had a significant impact in the field of multiferroics~\cite{hur_nature_2004,hur_prl_2004}.
Rare-earth multiferroic manganites, \RMn{} received much attention because of large ferroelectric polarization, puzzling magnetic structure and strong coupling between magnetic and ferroelectric orders~\cite{lee_prl_2013,hur_nature_2004,hur_prl_2004}, as compared to other multiferroics such as $R\text{MnO}_3$ \cite{goto_prl_2004,prokhnenko_prl_2007}, $\text{LiCu}_2\text{O}_2$ \cite{park_prl_2007}, and $\text{MnWO}_4$ \cite{taniguchi_prl_2006, niermann_prb_2014}. Because of the complex magnetic interactions and spin-lattice coupling, our current understanding of the underlying physics in \RMn{} still needs considerable efforts \cite{ratcliff_prb_2005,cruz_prb_2007}.

\subsection*{\label{RMn2O5}\RMn{}}

\begin{figure}
\centering
\includegraphics[width=0.6\linewidth]{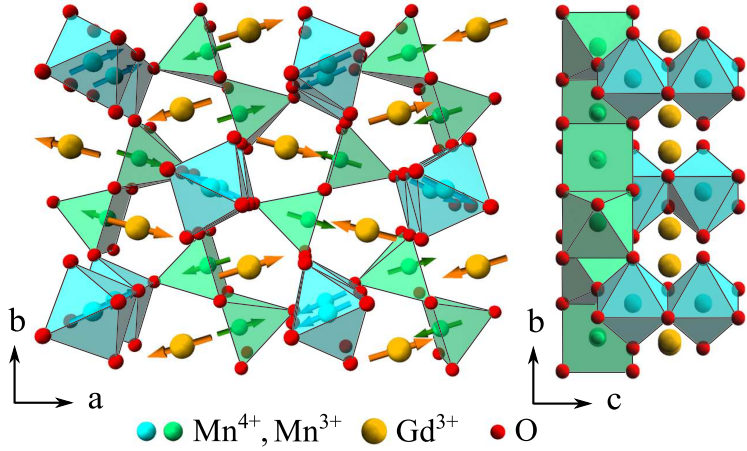}
\caption{Crystallographic structure of \GdMn{} in the commensurate magnetically-ordered state. Chains of light blue octahedra containing Mn$^{4+}$ ions run along the $c$ axis. Mn$^{3+}$ are located inside the green square-pyramids alternating with Gd$^{3+}$ in every second $a$-$b$ plane. Arrows indicate the spin orientations in the low-field low-temperature phase.}
\label{CrystStruct}
\end{figure}

\RMn{} has an orthorhombic crystal structure with \emph{Pbam} space group at room temperature\cite{bertaut_bullfr_1965, abrahams_jchph_1967} as exemplified in Fig.~\ref{CrystStruct} for \GdMn. The complex magnetism in this system originates from the $3d$ magnetic moments of Mn$^{3+}$, Mn$^{4+}$ and from the $4f$ magnetic moment of $R^{3+}$. There exist five relevant magnetic interactions between neighbouring spins, which makes the system highly frustrated and gives rise to various magnetically-ordered phases with long-period modulation wavelength~\cite{noda_jpcm_2008}. Due to inherent magnetic frustration  below the N\'{e}el temperature $T_{N1} \sim 40$~K, \RMn~shows a cascade of successive phase transitions upon temperature variation. In most members of the \RMn{} family, common phase transitions occur below $T_{N1}$, such as the high temperature incommensurate magnetic (HT-ICM) phase characterized by wave vector \textbf{q} = (q$_{x}$,0,q$_{z}$), and a lock-in transition from HT-ICM to the commensurate magnetic (CM) phase with \textbf{q} = (1/2,0,1/4), where ferroelectric polarization arises. On further cooling, the CM phase is destroyed and a new low temperature incommensurate (LT-ICM) phase may emerge. Moreover, in some cases, another phase appears between the HT-ICM and CM, known as the one-dimensional incommensurate (1D-ICM) phase \cite{noda_jpcm_2008}.

\subsection*{\label{GdMn2O5}\GdMn{}}

The magnetic structure of \GdMn{} is slightly different from other members of the \RMn~ family. Lee \emph{et al.} show~\cite{lee_prl_2013} that an incommensurate phase occurs below $T_{N1} \sim$~40 K with a wave vector $\mathbf{q} = (0.486,\, 0,\, 0.18)$, followed by a commensurate phase with $\mathbf{q} = (1/2,\, 0,\, 0)$ below $T_{N2} \sim 31$~K, which does not change further at low temperatures. This sequence of the wave vectors is due to the active involvement of Gd ions below $T_{C} \sim 29$~K because of their large atomic radius and high magnetic moments. This structure leads to tunable ferroelectricity in \GdMn{} in external magnetic fields~\cite{popov_phtt_2003, kadomtseva_ltp_2006}, inducing a giant change of polarization by $\sim 5000 \mu C$/$m^{2}$, the largest among the known multiferroic systems~\cite{lee_prl_2013}. Recently,
pressure-induced change in the polarization at the ferroelectric
transition in \GdMn~ has been observed and attributed to Gd-Mn
decoupling at high pressure due to lattice contraction~\cite{poudel_prb_2015, yin_jap_2016}.

In \RMn{} multiferroics, there exists strong coupling between electric polarization and magnetic order. The polarization can be induced by an applied magnetic field in $\text{HoMn}_2\text{O}_5$ and DyMn$_2$O$_5$ \cite{higashiyama_prb_2005, higashiyama_prb_2004, hur_prl_2004},  suppressed in $\text{ErMn}_2\text{O}_5$ \cite{higashiyama_prb_2005}, reversed in $\text{TbMn}_2\text{O}_5$ \cite{hur_nature_2004} and \GdMn{} \cite{lee_prl_2013} or flopped in $\text{TmMn}_2\text{O}_5$ \cite{fukunaga_prl_2009} and $\text{YbMn}_2\text{O}_5$ \cite{fukunaga_jpsj_2011}. These effects are observed in the low-temperature magnetically-ordered phases. Recently, the magnetic field effect on the electric polarization has been demonstrated in the NdMn$_2$O$_5$ multiferroic~\cite{chattopadhyay_prb_2016}.

In \RMn, the dielectric response in the external magnetic field depends strongly on the type of rare-earth and the direction of applied field. It has been found that magnetic fields affect the low temperature region (below $\sim 25$~K), leading to  remarkable magnetoelectric response. It indicates the key role of the $4f$ moments of the rare-earth ions in interaction with the external field which exists even at higher temperatures.

In this work, we present a set of dielectric properties $\varepsilon(H,T)$ along the crystallographic axes $a, b,$ and $c$ and in external magnetic fields. From the observed dielectric anomalies, we construct the magnetoelectric phase diagrams of \GdMn{}. Taking into account the results of previous studies, an assignment of various phases is suggested.

\section{\label{exp}Experimental}

Dielectric and magnetic properties of single crystalline \GdMn{} have been measured in magnetic fields up to 12~T and at temperatures between 2~K and 60~K using the Physical Property Measurement System. The system was supplemented by the vibrating magnetometer and by the Alpha impedance analyzer to measure the complex dielectric permittivity $\varepsilon$. Dielectric results shown originate from measurements at fixed frequency of 10 kHz.

The single crystal of \GdMn~ was grown by using the flux method \cite{lee_prl_2013, hur_nature_2004}, similar to other \RMn{} compounds. After determining the orientation of axes by Laue diffraction, it was cut into plane-parallel slices of area $\sim 5\times 5$~mm$^2$ and thickness $\sim$ 0.5~mm normal to the crystallographic $c$-axis. To obtain samples of other orientations, the slice dedicated for dielectric experiments was cut into several rectangular pieces parallel to the $a$ and $b$ axes. Dielectric experiments on samples with ac electric field $E \| a,b$ were performed on mosaic samples assembled from about 5 such fragments and fixed with G-varnish. Spontaneous electric polarization has been measured on a small crystal of about $0.4\times0.4\times0.4$mm$^3$ in size.  To obtain initial information on the characteristic temperatures and magnetic fields in \GdMn, additional dielectric experiments were carried out using polycrystalline samples.
Electric contacts to the samples were produced using silver paste.
The temperature scans of dielectric permittivity between 2 and 60 K were done under several fixed magnetic fields $\mu_0 H$, up to 12 T. Sweeping the magnetic field at selected temperatures provided additional information to complete the magnetoelectric phase diagrams.

In dielectric and magnetoelectric experiments, it is important to take care of possible artefacts due to conductivity~\cite{lunkenheimer_prb_2002} and magnetoresistance~\cite{catalan_apl_2006, schmidt_prb_2012} in the sample. In order to exclude such effects, several test experiments at different frequencies  were carried out. We found that the real part of the dielectric permittivity, $\varepsilon_1$, measured at frequencies of 1~kHz, 10~kHz, and 100~kHz agree to each other within few percent. In addition, the imaginary part of the permittivity, $\varepsilon_2$,  does not exceed $\sim 0.2$ in all experiments below $T=70$~K. We believe that these tests exclude extrinsic conducting effects in our measurements. Such effects could indeed be observed at temperatures above 100~K as $\sim 50 \%$ increase of $\varepsilon_1$ and as $\varepsilon_2$ increasing by factor of three.

\begin{figure*}
\centering
\includegraphics[width=\textwidth]{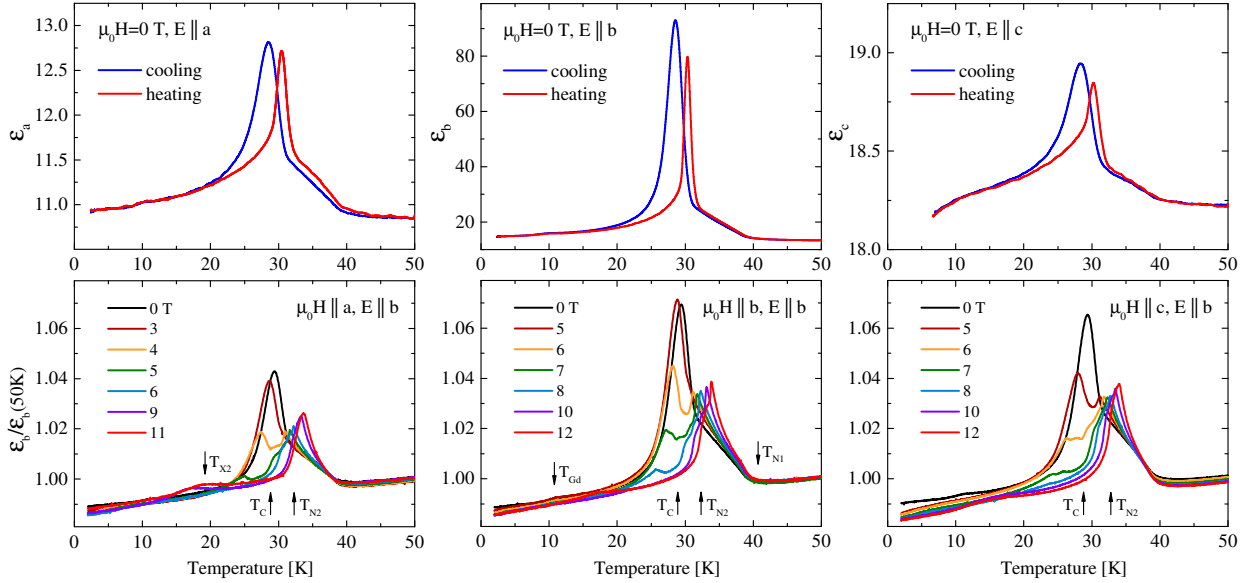}
\caption{Upper panels: Temperature dependent dielectric permittivity $\varepsilon_a$, $\varepsilon_b$ and $\varepsilon_c$ for \GdMn{} in zero magnetic field as obtained on a small single crystal. $E \| a,b,c$ is the \textit{ac} electric field.
Lower panels: Experiments in external magnetic field  up to $\mu_0 H = 12$~T along three crystallographic axes, as measured on a mosaic sample. The measurements were recorded on cooling run. The arrows indicate the temperatures of the phase transitions. }
\label{epsilon}
\end{figure*}

\section{\label{disc}Results}

\begin{figure}
\centering
\includegraphics[width=1.0\linewidth]{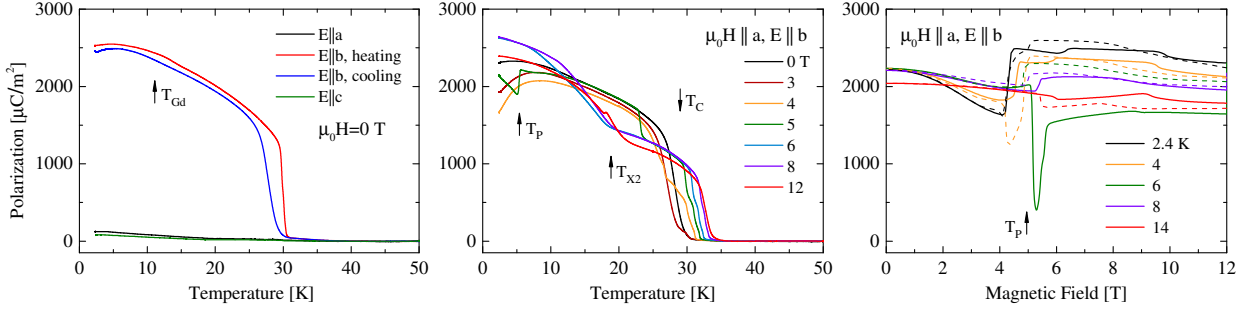}
\caption{Spontaneous electric polarization in \GdMn. Left panel: Polarization in zero magnetic field for three crystallographic directions as indicated. Middle panel: Electric polarization along the $b$-axis in the external magnetic field $H\| a$-axis. Right panel: Magnetic field-dependence of the polarization at different temperatures.  Positions of the phase transitions are marked by arrows.}
\label{fpolar}
\end{figure}

Fig.~\ref{epsilon} shows the temperature dependence of the dielectric permittivity along the $a, b,$ and $c$ axes in a range of magnetic fields (0-12 T), parallel to all three axes. The upper panels show the experiments in zero magnetic field on a small crystal. The lower panels show the results of the experiments in external magnetic field on a mosaic sample. The sample holder for external fields did not allow the absolute measurements due to a large stray signal. On cooling from the paramagnetic state, the dielectric permittivity starts to increase below $\sim 40$ K in zero magnetic field, which is the onset of long range incommensurate antiferromagnetic (ICM) order at $T_{N1}\sim$ 39 K. By further cooling, $\varepsilon$ exhibits a kink at $T_{N2} \sim 31 $~K and a peak at $T_{C} \sim 29 $~K where the  static polarization~\cite{lee_prl_2013,poudel_prb_2015} along the $b$-axis ($P \| b$) appears (see Fig.~\ref{fpolar}). This ferroelectric state for $t<T_{N2}$ coincides with  a commensurate magnetic order~\cite{lee_prl_2013} with $\mathbf{q} = (1/2,\, 0,\, 0)$.
In contrast to other \RMn, \GdMn~ reveals two close ferroelectric transitions at
$T_{N2}$ and $T_{C}$, respectively.

Near $T_C \sim 29$~ K, the dielectric permittivity exhibits a divergent behavior identifying another ferroelectric phase transition accompanied by a rapid growth of $P \| b$. Additionally, a small anomaly at low temperatures appears at $T \sim 11$~K. Since this temperature closely coincides with a maximum of the magnetic susceptibility along the $a$-axis (see Fig.~\ref{fMagnet}), we attribute this anomaly to Gd ordering transition~\cite{popov_phtt_2003, lee_prl_2013, tachibana_prb_2005}.

For several other members of the \RMn{} family, the main ferroelectric peak is nearly independent of the applied field~\cite{hur_prl_2004,higashiyama_prb_2005}. One of the interesting consequences of external magnetic fields in \GdMn~ is the transformation of the kink at $T_{N2}$ into a well pronounced peak which further moves towards higher temperatures with increasing magnetic field $\mu_0 H_a >$ 3 T. In addition, the sharp peak at $T_C$ decreases rapidly and moves towards lower temperatures with increasing $H_a$ and fully vanishes at $\mu_0 H_a \sim$ 6 T.  In the geometry $H \| a$, a new peak appears in the low temperature range  at $T_{X2} \sim$ 20 K, above 6 T. This temperature separates two ferroelectric phases assigned as FE1 and FE5 (see Fig.~\ref{PhD}).

Applying magnetic fields along the $b$ and $c$ axes, qualitatively similar dependencies are observed (Fig.~\ref{epsilon}), except for substantially higher absolute values of the characteristic fields at the phase transitions. This agrees well with the fact that the $a$-axis is the easy magnetic axis.

The results of the static polarization experiments on a small \GdMn~ crystal are shown in Fig.~\ref{fpolar}. The left panel of Fig.~\ref{fpolar} shows the data for the zero magnetic field and it demonstrates that static polarization in \GdMn~can be seen solely along the $b$-axis in the zero magnetic field. We note that spontaneous polarization shows a well-defined step and substantial hysteresis at $T_C$. Both features are indications of the first-order transition in \GdMn. As may be expected from the results of dielectric experiments, several additional phase transitions are observed in electric polarization of \GdMn~in external magnetic fields. The splitting of the ferroelectric transition at $T_C \sim 29$~K is visible as complex behavior of the polarization around this temperature. In high magnetic fields (middle panel) $\mu_0 H \geq 6 T$, the saturation tendency of $P_b \rightarrow 1500 \mu C/cm^2$ below $T_C$ is interrupted close to $T_{X2} \sim 20$~K and an additional contribution to the polarization starts to grow. As a result, the low-temperature polarization in \GdMn~ reaches $P_b \sim 2500 \mu C/cm^2$. As will be discussed below, we attribute an additional contribution to the influence of Mn-Gd interactions.

Another interesting result of the polarization experiments is the appearance of a sharp step-like feature in $P_b$ at $T_P \sim$ 5 K in the field range of 4.5-5.5 T which is marked by arrows in the middle and right panels of Fig.~\ref{fpolar}. This feature is accompanied by a peak in the dielectric permittivity at the same temperature. The behavior of electric polarization around $T_P$ may be a hint for a possible partial polarization flop from the $P \| b$ axis to the $a$ or $c$ axes. Similar polarization flop occurring at $\sim 5$~K in external magnetic fields have been observed earlier in  $\text{YbMn}_2\text{O}_5$~[\onlinecite{fukunaga_jpsj_2011}] and in TmMn$_2$O$_5$~[\onlinecite{fukunaga_prl_2009}]. However, in the present experiments no spontaneous polarization along the $a$ or $c$ axes could be observed within the experimental accuracy.

\subsection*{\label{MEPhD}Magnetoelectric phase diagrams}

\begin{figure}
\centering
\includegraphics[width=0.5\linewidth]{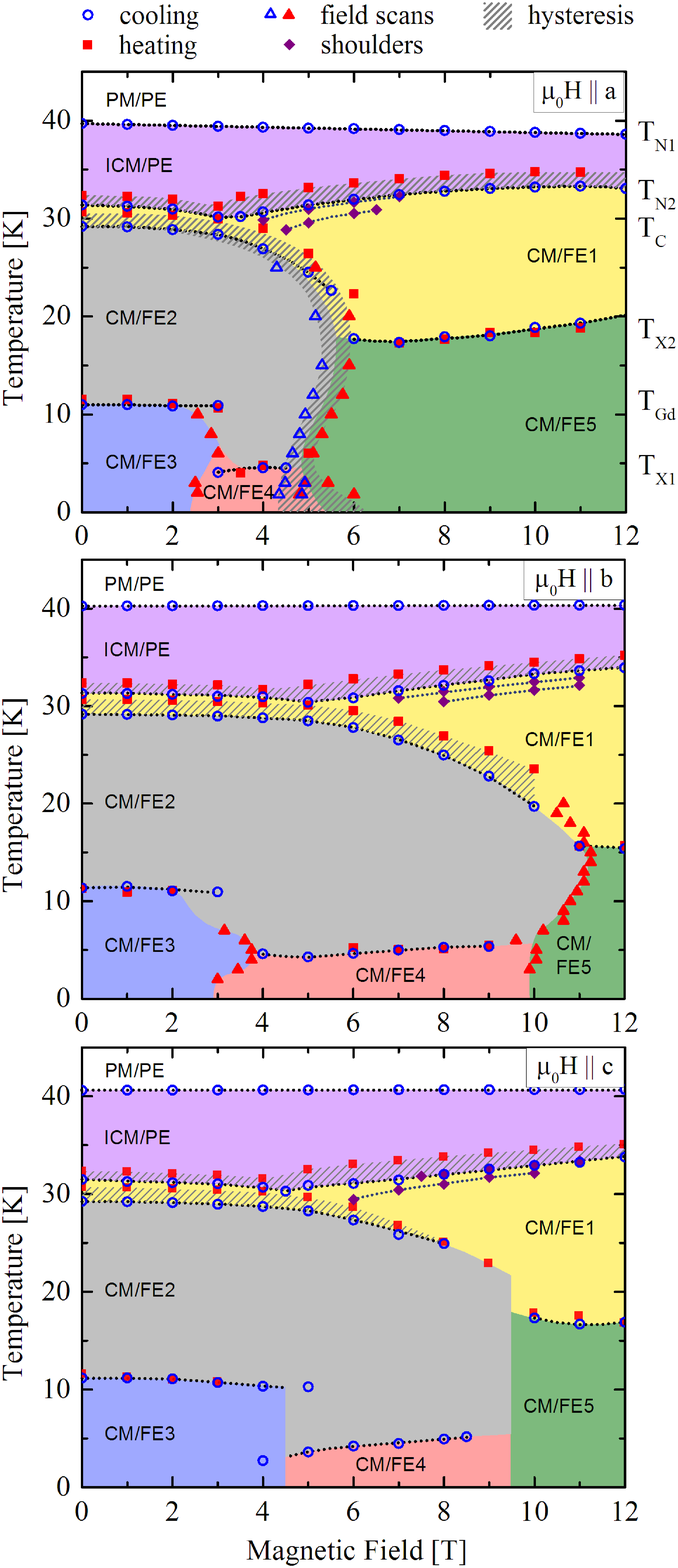}
\caption{Magnetoelectric phase diagrams of \GdMn{} with magnetic field along the $a, b,$ and $c$ axes. Open circles and closed squares represent the data obtained with decreasing and increasing temperature, respectively. Open and closed triangles denote the data obtained in magnetic field sweep in increasing and decreasing fields, respectively. Shaded regions show the hysteresis effects.}
\label{PhD}
\end{figure}

\begin{figure*}
\centering
\includegraphics[width=0.99\linewidth]{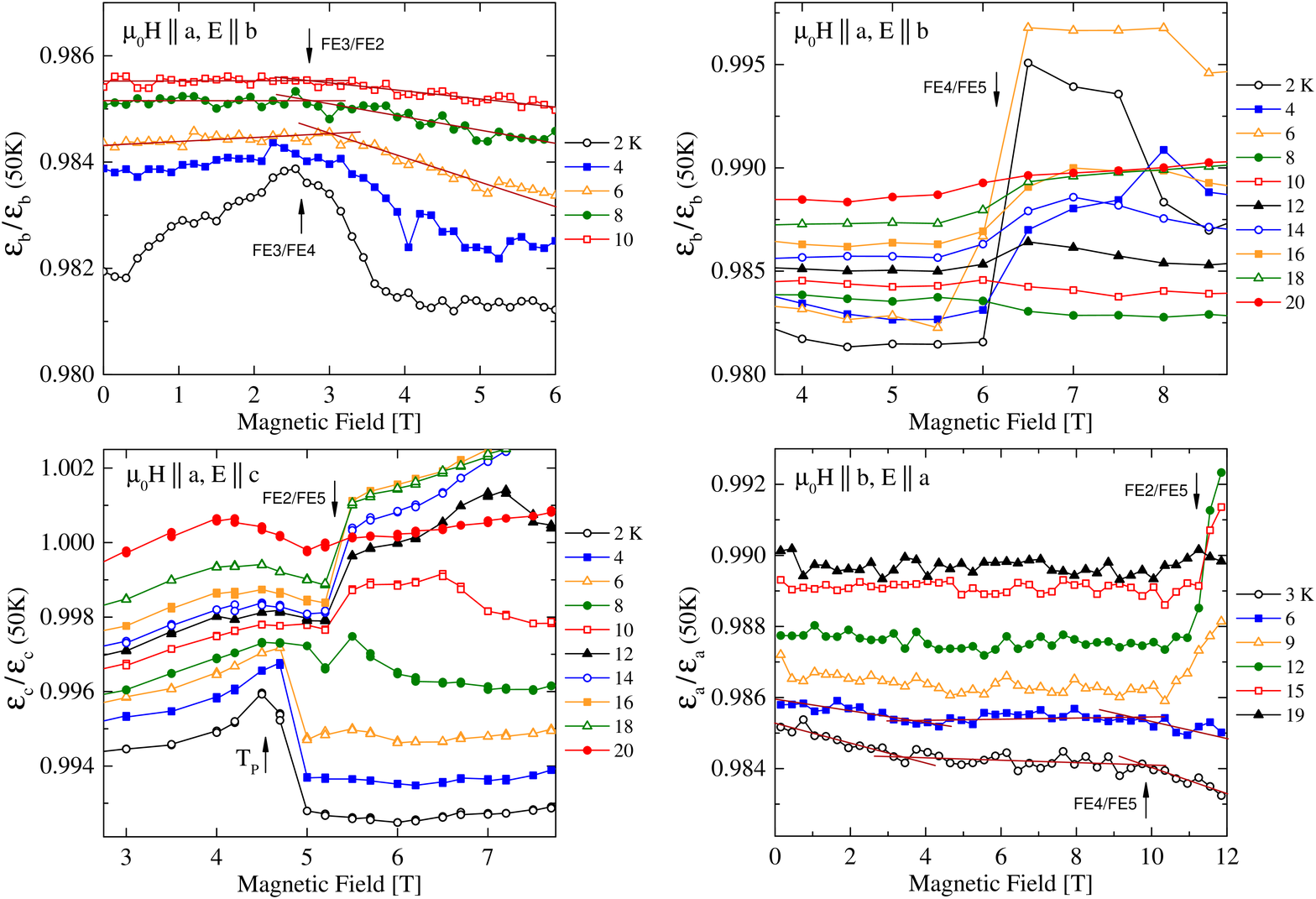}
\caption{Example of magnetic field-dependence of dielectric permittivity in \GdMn{} at various temperatures. Straight red lines show the position of changes-in-slope in $\varepsilon(H)$. Arrows indicate the suggested field values at the phase boundaries.}
\label{fieldscan}
\end{figure*}

Summarizing our experimental results on electric and magnetic properties of \GdMn, we show the magnetoelectric phase
diagrams for external magnetic fields along the $a$, $b$, and $c$ axes, as shown
in Fig. \ref{PhD}. The two phase boundaries defined by ${T_{N1}}$ and
${T_{N2}}$ in the phase diagram, are similar for external magnetic fields along all axes.
These phase boundaries correspond to the phase transitions PM/PE
$\rightarrow$ ICM/PE at ${T_{N1}}$ and ICM/PE $\rightarrow$ CM/FE1
at ${T_{N2}}$. The phase boundaries  of
${T_{N2}}$ and ${T_{C}}$ show a strong hysteresis effect indicating
the first-order transition~\cite{higashiyama_prb_2005}.

To support the determination of the phase boundaries, magnetic field scans of the dielectric permittivity $\varepsilon(H)$ below 40 K and in the fields up to $\mu_0 H = 12$~T were performed and are shown in Fig.~\ref{fieldscan}. For $H \| a$ and $E \| b$ (top left), in varying fields and at low temperatures $T \leq$ 10 K, an anomaly has been identified around 3 T, which  coincides with the phase boundary drawn from temperature scans. With increasing fields up to 6 T (top right), a clear jump-like anomaly is evident below $\sim 18$~K, which coincides with the Gd spin-flop transition~\cite{lee_prl_2013}. This phase boundary is only weakly temperature-dependent and, therefore, it has not been observed in fixed-field scans. For $H \| b$ and $E \| a$ (bottom right), strong anomalies of $\varepsilon_a (H)$ around 11~T have been detected, which corresponds to the same phase boundary in the $H \| b$ geometry.

In the case of $H \| a$, the boundary at $T_C$ between the FE2 and FE1 phases smoothly transforms to the FE2/FE5 phase boundary around 5.5 T. At low temperatures ($T < 6$~K), additional structure is seen in several magnetic field-dependent experiments. Because this temperature range corresponds to an ordered state of the Gd spins, a possible explanation for additional boundaries may be a reorientation of the Gd sublattice. As mentioned above, a clear peak in $\varepsilon_a$ has been observed at this boundary.

\begin{figure*}
\centering
\includegraphics[width=0.8\linewidth]{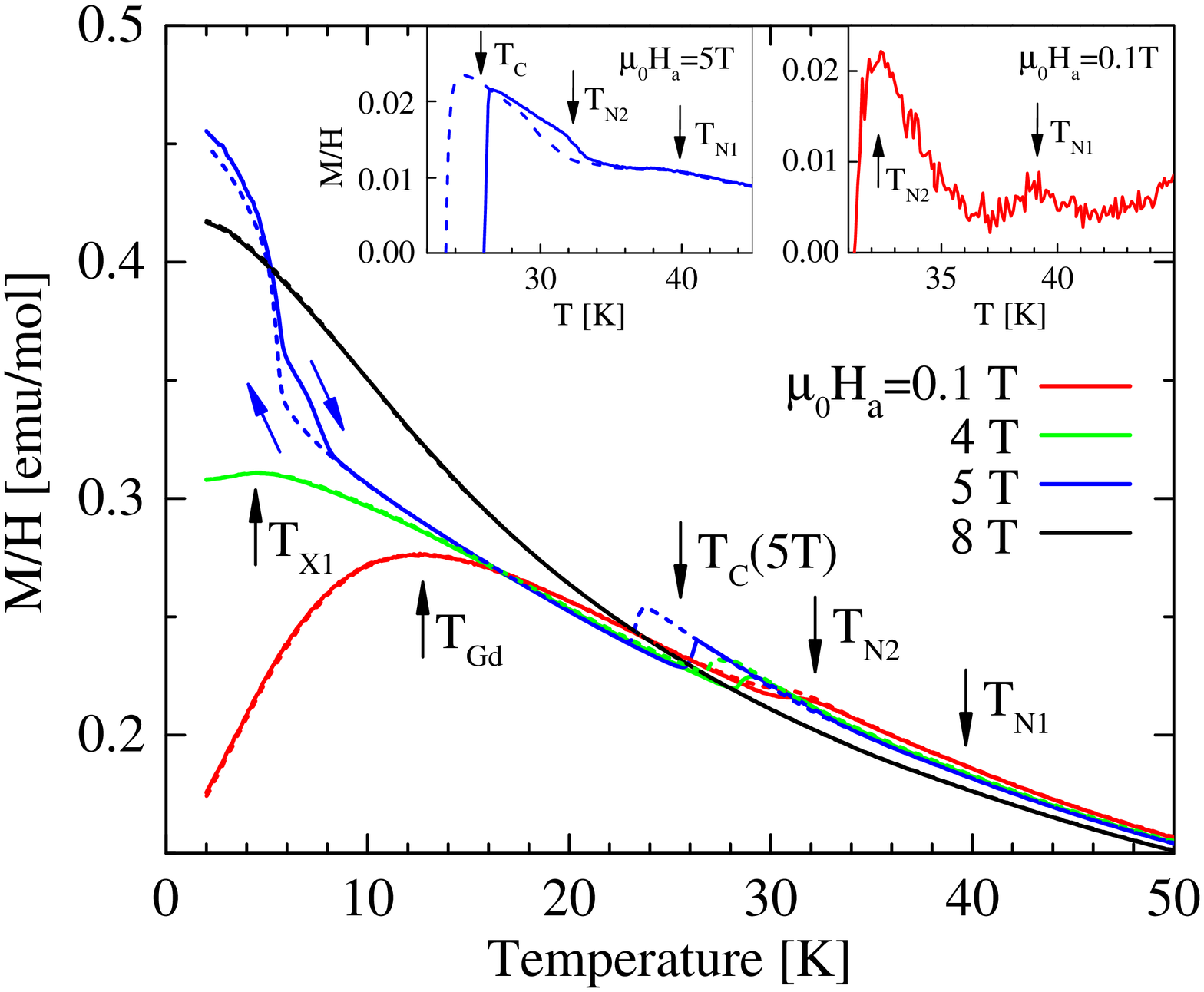}
\caption{Temperature dependence of the magnetization in \GdMn{} along the easy axis and at several external fields, as indicated. All main phase transitions are well seen in the magnetic data. Left and right insets show the magnetization at $\mu_0 H_a = 5$~T and at $\mu_0 H_a = 0.1$~T, respectively,  with smooth background subtracted to reveal the magnetic transition temperatures $T_{N1}$ and $T_{N2}$. Arrows indicate the transition temperatures between different magnetic phases.}
\label{fMagnet}
\end{figure*}

We recall that in magnetic field-dependent experiments (Fig.~\ref{fieldscan}), several step-like features have been observed in \GdMn. According to the sum rule for the static dielectric permittivity~\cite{dressel_book_2002}, an electrically dipole-active excitation may then be suggested to exist at high frequencies. Similar to electromagnons in several multiferroics \cite{shuvaev_prl_2010, sushkov_prb_2014}, a step in the dielectric permittivity may directly correspond to a change in dielectric contribution of a high-frequency mode.

In order to complete the magnetoelectric phase diagram, magnetization measurements in \GdMn~ have been carried out in several magnetic fields along the $a$-axis, as shown in Fig.~\ref{fMagnet}. The paramagnetic susceptibility above $T_{N1} \sim 39$~K approximately follows the Curie-Weiss law $H/M \sim (T+\theta)$ with $\theta$ around 25~K. The N\'{e}el transition temperature is seen in the magnetization data after subtraction of a smooth function of temperature, which is shown in the insets of Fig.~\ref{fMagnet}. One prominent feature in the low-field magnetization is a broad maximum around $T_{Gd} \sim 10$~K which is attributed to the ordering of the Gd subsystem. A step-like feature at $\mu_0 H = 5$~T and around 8~K probably corresponds to the spin-flop transition of Gd-ions. This transition is clearly seen in the field-dependent magnetization curves until $T=30$~K (not shown), suggesting the existence of Gd-order in both high-field phases, FE1 and FE5.

The phase diagrams for $H \| b$ and $H \| c$ exhibit similar phase sequences as observed for $H \| a$, despite of the difference in absolute values of critical fields. Therefore, we suggest that fundamental physics in the different phases (FE1 $\rightarrow$ FE5) is the same in all \GdMn~ phase diagrams.

\section{\label{Shift}Discussion}

\begin{figure}
\centering
\includegraphics[width=0.75\linewidth]{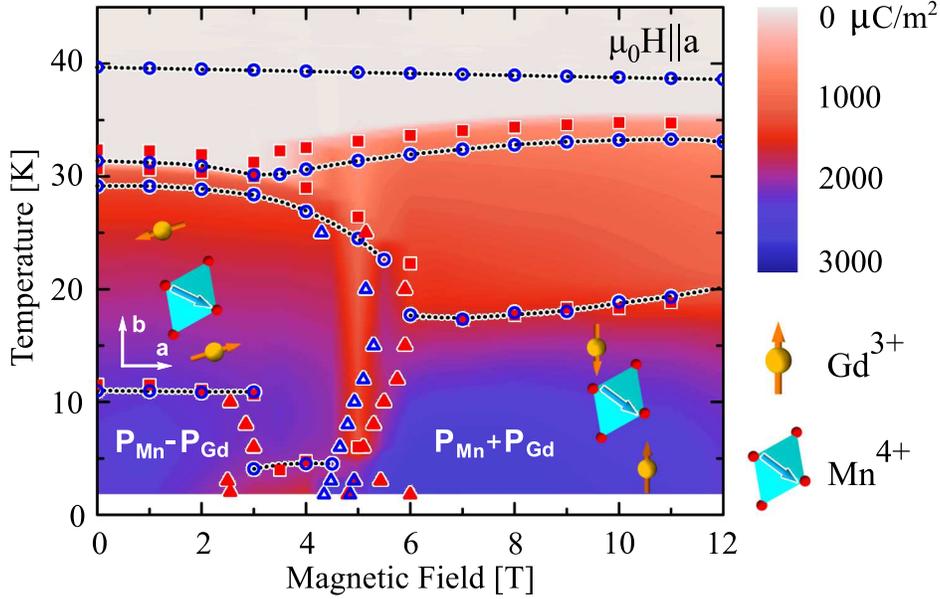}
\caption{Color-coded magnetoelectric phase diagram of \GdMn~in the $H \|a$ geometry, as obtained from the interpolation of the polarization data of Fig.~\ref{fpolar}. Relative orientation of Mn and Gd moments are given as pictograms illustrating the suggested $90^\circ$ rotation of Gd spins.}
\label{polaris}
\end{figure}

The series of the magnetoelectric transitions presented here is in agreement with previous experiments in \GdMn~[\onlinecite{lee_prl_2013, poudel_prb_2015, yin_jap_2016}]. Figure~\ref{polaris} reproduces the main features of the phase diagram in Fig.~\ref{PhD} (for $H \|a$) and of the field-dependent polarization given in Fig.~~\ref{fpolar}. The polarization data are given as color scale and they fit well in the magnetoelectric phase diagram.

Following the arguments of Lee \emph{et al.}, Ref.~[\onlinecite{lee_prl_2013}], the boundary around 5.5 T between the FE2 and FE5 phases corresponds to a switching of additional mechanisms of electric polarization in \GdMn. The total electric polarization may be obtained as a competition of two contributions originating from the symmetric exchange-striction mechanism in Mn-Mn and Mn-Gd spin-subsystems, respectively. In magnetic fields exceeding $\mu_0 H_a \sim 5.5$~T, the Gd spins rotate by $90^\circ$ leading to the inversion of the Mn-Gd contribution to electric polarization. As a result, the total polarization in the right panel of  Fig.~\ref{fpolar} increases from $P_b(0~\mathrm{T})$ (FE2,3)~$\sim 1500$~$\mu C/m^2$ to about $P_b(8~\mathrm{T}) \sim 2500$~$\mu C/m^2$.

Magnetic field dependence of the electric polarization in Fig.~\ref{fpolar} is rather complex. Remarkable is a sharp decrease of $P_b$ in at fields $\sim 5$~T, which is followed by a step-like increase for further increasing fields. Conversely, in previous  experiments~\cite{lee_prl_2013} only a decrease of electric polarization has been observed reaching even negative values and remaining stable until $B=9$~T. Most probably, the complex magnetic structure of \GdMn~is sensitive to the exact orientation of the external field and to details of cooling and poling procedures. Further investigations are necessary to resolve this problem.


\section{\label{concl}Conclusions}

Magnetodielectric effects for single crystalline \GdMn{} have been investigated with dielectric spectroscopy, electric polarizations, and magnetic experiments. Several phase transitions are observed which are strongly field-dependent with respect to field orientation and strength. A full set of magnetoelectric phase diagrams in external magnetic fields has been determined. Qualitative similarity of the phase diagrams suggests that the same sequence of magnetoelectric phases exists for all geometries in \GdMn.

\subsection*{\label{ackn}Acknowledgements}

S. H. Bukhari acknowledges financial support by the Higher Education Commission (HEC) of Pakistan through a IRSIP scholarship. This work was supported by the Austrian Science Funds (I815-N16, I1648-N27, W1243). The work at Rutgers University was supported by the DOE under Grant No. DOE: DE-FG02-07ER46382.

\bibliography{literature}

\end{document}